\documentclass[12pt]{article}
\usepackage{latexsym}
\usepackage{amsmath}
\usepackage{cite} 
\usepackage{indentfirst}
\usepackage{latexsym}
\usepackage{pstricks}

\long\def\ca#1\cb{} 

\newcommand{\ad}{^\dagger }

\newcommand{\AND}{\mbox{\small AND}}

\newcommand{\becs}{\begin{cases}}
\newcommand{\bem}{\begin{matrix}}

\newcommand{\dya}[1]{|#1\rangle\langle#1|}

\newcommand{\encs}{\end{cases}}
\newcommand{\enm}{\end{matrix}}

\newcommand{\inpd}[2]{\langle#1|#2\rangle }
\newcommand{\ket}[1]{|#1\rangle }

\newcommand{\mte}[2]{\langle#1|#2|#1\rangle }

\newcommand{\NOT}{\mbox{\small NOT}}

\newcommand{\OR}{\mbox{\small OR}}

\newcommand{\gm}{\gamma }
\newcommand{\Gm}{\Gamma }
\newcommand{\dl}{\delta }

\newcommand{\lm}{\lambda }

\begin{document}

\title{Quantum Information: What Is It All About?}
\author{Robert B. Griffiths\thanks{Electronic address: rgrif@cmu.edu}\\
Department of Physics\\
Carnegie Mellon University\\
Pittsburgh, PA 15213}
\date{Version of 28 November 2017}
\vspace*{-2cm}

{\let\newpage\relax\maketitle}

\vspace{-1cm}

\begin{abstract}
  This paper answers Bell's question: What does quantum information refer to?
  It is about quantum properties represented by subspaces of the quantum
  Hilbert space, or their projectors, to which standard (Kolmogorov)
  probabilities can be assigned by using a projective decomposition of the
  identity (PDI or framework) as a quantum sample space. The single framework
  rule of consistent histories prevents paradoxes or contradictions. When only
  one framework is employed, classical (Shannon) information theory can be
  imported unchanged into the quantum domain. A particular case is the
  macroscopic world of classical physics whose quantum description needs only a
  single quasiclassical framework. Nontrivial issues unique to quantum
  information, those with no classical analog, arise when aspects of two or
  more incompatible frameworks are compared.
\end{abstract}

\tableofcontents

\section{Introduction \label{sct1}}

A serious study of the relationship between quantum information and
quantum foundations needs to address Bell's rather disparaging question,
``Quantum information ... about what?'' found in the third section 
of his polemic against the role of measurement in standard (textbook) quantum
mechanics \cite{Bll901}. The basic issue has to do with quantum ontology,
``beables'' in Bell's language. I believe a satisfactory answer to Bell's
question is available, indeed was already available (in a somewhat preliminary
form) at the time he was writing. (If he was aware of it, Bell did not mention
it in any of his publications.) Further developments have occurred since, and I
have found this approach to be of some value in addressing some of the
foundational issues which have come up during my own research on quantum
information. So I hope the remarks which follow may assist others who find the
textbook (both quantum and quantum information) presentations confusing or
inadequate, and are looking for something better.

Here is a summary of the remainder of this paper. The discussion begins in
Sec.~\ref{sct2} by asking Bell's question about \emph{classical} (Shannon)
information: what is \emph{it} all about? That theory works very well in the
world of macroscopic objects and properties. Hence if classical physics is
fundamentally quantum mechanical, as I and many others believe, and if Shannon's
approach is, as a consequence, quantum information theory applied to the domain
of macroscopic phenomena, we are already half way to answering Bell's question.
The other half requires extending Shannon's ideas into the microscopic domain
where classical physics fails and quantum theory is essential. This is
possible, Sec.~\ref{sct3}, using a consistent formulation of standard
(Kolmogorov) probability theory applied to the quantum domain. Current quantum
textbooks do not provide this, though their discussion of measurements,
Sec.~\ref{sct4}, gives some useful hints. The basic approach in Sec.~\ref{sct3}
follows von Neumann: Hilbert subspaces, or their projectors, represent quantum
properties, and a projective decompositions of the identity (PDI) provides a
quantum sample space. By not following Birkhoff and von Neumman, but instead
using a simplified form of quantum logic, Sec.~\ref{sct5}, one has, in the
``single framework rule'' of consistent histories, a means of escaping the
well-known paradoxes that inhabit the quantum foundations swamp.
Section~\ref{sct6} argues that when quantum theory is equipped with (standard!)
probabilities, quantum information theory is identical to Shannon's theory in
the domain of macroscopic (classical) physics, as one might have expected,
since only a single quasiclassical quantum framework (PDI) is needed for a
quantum mechanical description. However, classical information theory also
applies, unchanged, in the microscopic quantum domain if only a \emph{single
  framework} is needed. Section~\ref{sct7} provides a perspective on the highly
nontrivial problems that are unique to quantum information and lack any simple
classical analog: they arise when one wants to \emph{compare} (not combine!)
two or more \emph{incompatible} frameworks applied to a particular
situation.

\section{Classical Information Theory \label{sct2}}

Let us start by asking Bell's question about \emph{classical} information
theory, the discipline which Shannon started. What's it all about? If you open
any book on the subject you will soon learn that it is all about probabilities,
and information measures expressed in terms of probabilities. So we need to
ask: probabilities of what? Standard (Kolmogorov) probability theory, the sort
employed in classical information theory, begins with a \emph{sample space} of
mutually exclusive possibilities, like the six faces of a die. Next an
\emph{event algebra} made up of subsets of elements from the sample space, to
which one assigns \emph{probabilities}, nonnegative numbers between 0 and 1,
satisfying certain \emph{additivity} conditions.

The simplest situation, quite adequate for the following discussion, is a
sample space with a finite number $n$ of mutually exclusive possibilities, let
them be labeled with an index $j$ between $1$ and $n$ (or $0$ and $n-1$ if
you're a computer scientist). The event algebra consists of all $2^n$ subsets
(including the empty set) of elements from this sample space. Then for
probabilities choose a collection of $n$ nonzero real numbers $p_j$ lying
between 0 and 1, which sum to 1. The probability of an element $S$ in the event
algebra is the sum of the $p_j$ for $j$ in $S$.

The mutually exclusive possibilities might be distinct letters of an alphabet
used to send messages through a communication channel, and in the actual
physical world each letter will be represented by some unique \emph{physical
  property(s)} that identifies it and distinguishes it from the other letters
of the alphabet. One way to visualize this is to think of a classical phase
space $\Gm$ in which each point $\gm$ represents the precise state of a
mechanical system, and a particular letter of the alphabet, say $F$, is
represented by some collection of points in $\Gm$, the set of points where the
property corresponding to $F$ is true, and where the corresponding
\emph{indicator function} $F(\gm)$ takes the value 1, whereas for all other
$\gm$, $F(\gm)=0$. The different indicator functions associated with letters of
the alphabet then split the phase space up into tiles, regions in which a
particular indicator function for a particular letter is equal to 1, and
indicators for the other letters are all equal to zero. If this tiling does not
cover the entire phase space, simply add another letter to the alphabet, call
it ``NONE'', and let its indicator be 1 on the remaining points, and 0
elsewhere. In this manner one can map the abstract notion of an alphabet of
mutually exclusive letters onto a collection of mutually exclusive physical
properties, one and only one of which will be true at any given time, because
the point in phase space representing the actual state of the mechanical system
will be located in just one of the nonoverlapping tiles. Given the sample space
of tiles and some way of assigning probabilities, we have a setup to which the
ideas of classical information theory can be applied, with a fairly clear
answer to the question of what the information is all about.

In summary, classical information theory is all about probabilities, and in any
specific application, say to signals coming over an optical fiber, the
probabilities are about, or make reference to, physical events or properties
of physical systems.

\section{Quantum Probabilities \label{sct3}}

If we want quantum information theory to look something like Shannon's theory,
the first task is to identify a quantum sample spaces of mutually-exclusive
properties to which probabilities can be assigned. The task will be simplest if
these quantum probabilities obey the same rules as their classical
counterparts. In particular, since Shannon's theory employs expressions like
$p_j\log(p_j)$, it would be nice if the quantum probabilities were nonnegative
real numbers, in contrast to the negative quasiprobabilities sometimes
encountered in discussions of quantum foundations.

Can we identify a plausible sample space which relative to the quantum Hilbert
space plays a similar role to a tiling of a classical phase space? (In what
follows I will assume that the quantum Hilbert space is a finite-dimensional
complex vector space with an inner product. Thus all subspaces are closed, and
we can ignore certain mathematical subtleties needed for a precise discussion
of infinite-dimensional spaces.) A useful beginning is suggested by the
quantum textbook approach to probabilities given by the Born rule. Let $A$ be an
observable, a Hermitian operator on the quantum Hilbert space, and let
\begin{equation}
 A = \sum_j a_j P_j
\label{eqn1}
\end{equation}
be its spectral representation: the $a_j$ are its eigenvalues and the $P_j$
are projectors, orthogonal projection operators, which form a 
\emph{projective decomposition of the identity $I$} (PDI):
\begin{equation}
 I = \sum_j P_j; \quad P^{}_j = P_j\ad;\quad P_j P_k = \dl_{jk} P_j.
\label{eqn2}
\end{equation}
If the eigenvalue $a_j$ is nondegenerate and $\ket{\phi_j}$ is the corresponding
eigenvector, then 
\begin{equation}
  P_j = \dya{\phi_j} = [\phi_j],
\label{eqn3}
\end{equation}
where $[\phi]$ is a convenient abbreviation for the
Dirac dyad $\dya{\phi}$.

According to the textbooks, given a normalized ket $\ket{\psi}$, the probability
that when $A$ is measured the outcome is $a_j$, is given by the Born rule:
\begin{equation}
 p_j=\Pr(a_j) = \Pr(P_j) = \mte{\psi}{P_j} = |\inpd{\psi}{\phi_j}|^2,
\label{eqn4}
\end{equation}
where the final equality applies only when $P_j$ is the rank one projector in
\eqref{eqn3}. Now a measurement of $A$ will yield just one eigenvalue, not
many, so these eigenvalues correspond to the mutually-exclusive properties
$P_j$ in the PDI used in \eqref{eqn4}. The idea that a quantum property should
be associated with a subspace of the Hilbert space, or the corresponding
projector, goes back at least to von Neumann, see Sec.~III.5 of his oft-cited
(but little read) book \cite{vNmn55}.

The projector $P_j$ has eigenvalues 0 and 1, so it resembles an indicator
function on the classical phase space. In fact, a PDI divides up the Hilbert
space into a set of mutually exclusive subspaces---$P_jP_k=0$ for $j\neq
k$---somewhat like a tiling of the classical phase space, whereas $I = \sum_j
P_j$ tells us this tiling is complete: no part of the Hilbert space has been
left out. Thus the PDI is a plausible candidate for a quantum sample
space. The event algebra will then consist of the projectors in the PDI along
with other projectors formed from their sums, including $I$, along with the
zero operator. The result is a commutative Boolean algebra. We already have one
scheme, \eqref{eqn4}, for assigning probabilities to elements of the PDI, and
thus, by additivity, to all the projectors in the event algebra. In
particular, for $j\neq k$,
\begin{equation}
  \Pr(P_j\ \OR\ P_k) = \Pr(P_j) + \Pr(P_k) = \mte{\psi}{(P_j+ P_k)},
\label{eqn5}
\end{equation}
and similarly for sums of three or more distinct projectors.  

In summary, this looks like a plausible beginning for a theory of \emph{quantum
  information}: use a PDI on the Hilbert space as a sample space; then assign
probabilities to the individual projectors. Not necessarily using \eqref{eqn4},
for it is only a particular example, but by some scheme which yields
nonnegative real numbers adding to 1. Indeed, this strategy works very well,
and I believe it covers all legitimate uses of (standard) probability theory in
quantum mechanics, at least for a Hilbert space of finite dimension.

\section{Quantum Measurements \label{sct4}}

There is, of course, more to be said, and it can be motivated by noting that a
carefully written quantum textbook is likely to assign the probability $p_j$
not to the \emph{microscopic property} of the measured system, represented by
$P_j$, but instead to the \emph{macroscopic measurement outcome}, the
\emph{pointer position} in the picturesque, albeit archaic, language of quantum
foundations. But in the above presentation it looks as if the probability
is assigned \emph{directly} to the microscopic property. Was this a mistake?
Not if one believes, as I do, that a properly constructed and calibrated
apparatus designed to measure some quantum observable can actually do what it
was designed to do. And if there is a one-to-one correspondence between prior
properties and later pointer positions, the probability $p_j$ will be the same
for both.

In support of my belief that quantum measurements measure something, I note
that this is assumed by my colleagues who do experiments at accelerator
laboratories. They think that when they detect a fast muon emerging from an
energetic collision, there really was a fast muon that approached and triggered
their detector. Are they being naive? I don't think so. And in passing I note
that these colleagues don't seem to worry about the ``collapse'' of the muon
wavefunction produced by its interaction with the detector; they are less
interested in what happened to the muon after it left their measuring device,
and more interested in knowing what it was doing before it arrived there.

In addition, the notion that outcome $j$ corresponds to the earlier property
$P_j$ can in certain cases be tested by \emph{preparing} a particle which has
the property $P_j$ (see Sec.~IV~C of \cite{Grff17b} on the topic of
preparation), sending it into the measurement apparatus, and seeing whether the
result is that the pointer points to $j$. Given that the apparatus has been
tested and calibrated in this way, is not the experimenter justified in
thinking that the particle \emph{had} the property indicated by the pointer in
a run in which the particle was \emph{not} prepared in one of the $P_j$ states?
Justified or not, this is how many of my colleagues who carry out experiments
do interpret things, and if they didn't it would be difficult to draw
interesting conclusions from their data. Quantum physics can hardly be called
an \emph{experimental} science if experiments designed to reveal prior
microscopic properties do not actually do so! For additional details on the
topic of what quantum measurements measure, including POVM and weak
measurements, see \cite{Grff17b}.

There is, to be sure, a conceptual difficulty lurking in the background if we
assume that measurements reveal prior microscopic properties. A hint is
provided by the (correct) statement in textbooks that the $x$ and $z$
components of spin angular momentum, $S_x$ and $S_z$, of a spin-half particle
\emph{cannot be measured simultaneously}. True, but what principle lies behind
this? If we assume that experimenters really do understand something about what
their devices measure, their inability to carry out such a simultaneous
measurement might plausibly be explained by the fact that \emph{there is
  nothing there to be measured.} Even very skilled experimenters cannot measure
what isn't there; indeed, this could be one thing that distinguishes them from
less capable colleagues.

The Hilbert space of a spin-half particle is two-dimensional, and while it
contains two subspaces corresponding to $S_x=\pm 1/2$ (in units of $\hbar$),
and another two corresponding to $S_z=\pm 1/2$, there is no subspace which can
plausibly be associated with, to take an example, ``$S_x=+1/2$ \AND\
$S_z=-1/2$''. Hence if we assume that quantum measurements measure microscopic
properties represented by subspaces of the quantum Hilbert space (or their
projectors), we have a ready explanation for what lies behind the assertion
that $S_x$ and $S_z$ cannot both be measured simultaneously. This is
one way in which quantum mechanics is very different from classical mechanics.

\section{Incompatible Properties \label{sct5}} 

\subsection{Issues of Logic \label{sbct5.1}}

The absence of a Hilbert subspace corresponding to ``$S_x=+1/2$ \AND\
$S_z=-1/2$'' reflects an important difference between the logic of
indicator functions on the classical phase space and quantum projectors on the
Hilbert space. One analogy has already been noted: the indicator $F(\gm)$ for a
classical property $F$ takes one of two values, 0 and 1, while a quantum
projector $P$ has eigenvalues that are either 0 or 1. In addition, the negation
``\NOT\ $F$'' of a classical property has an indicator function $I(\gm) -
F(\gm)$, where $I(\gm)$ is the function which is equal to 1 everywhere on
the phase space. Similarly, the negation ``\NOT\ $P$'' of a quantum projector
$P$ is the projector $I-P$, with $I$ the quantum identity operator. But the
analogy begins to break down when we consider the conjunction ``$F$ \AND\ $G$''
of two classical properties: the property which is true if and only if both $F$
and $G$ are true. It corresponds to the intersection of the two subsets of
phase space points associated with $F$ and $G$, and its indicator is the
product $F(\gm)G(\gm)$ of the two indicators. So we might expect that the
conjunction ``$P$ \AND\ $Q$'' of two quantum properties $P$ and $Q$ would be
represented by the product $PQ$. Indeed, this is the case \emph{if} the
projectors $P$ and $Q$ \emph{commute}, $PQ=QP$, in which case $PQ$ is again a
projector. However, if $PQ$ is \emph{not} equal to $QP$, then neither product
is a projector, and it is not obvious how to define ``$P$ \AND\ $Q$''.

The point can be illustrated using $S_x$ and $S_z$ for a spin-half particle.
The projectors representing $S_x=+1/2$ and $-1/2$ are $[x^+] = \dya{x^+}$ and
$[x^-]$, where $\ket{x^+}$ and $\ket{x^-}$ are the eigenvectors corresponding
to $S_x=+1/2$ and $-1/2$. Since $\inpd{x^+}{x^-}=0$ (distinct eigenvalues means
the eigenvectors are orthogonal) $[x^+][x^-] = [x^-][x^+]=0$. Thus these
projectors commute, and the property ``$S_x=+1/2$ \AND\ $S_x=-1/2$'' is
represented by the zero operator on the Hilbert space: the property that is
always false and thus never occurs. Also $[x^+]+[x^-]=I$ so these two
mutually-exclusive properties constitute a PDI, a quantum sample space. Likewise
the projectors $[z^+]$ and $[z^-]$ that correspond to
$S_z=+1/2$ and $-1/2$ form a PDI.

However, neither $[x^+]$ nor $[x^-]$ commutes with either $[z^+]$ or $[z^-]$,
so we cannot assign a quantum property to ``$S_x=+1/2$ \AND\ $S_z=-1/2$'' by
taking the product of the projectors. Again, this is consistent with the idea
that the reason a simultaneous measurement of $S_x$ and $S_z$ is impossible is
that there is nothing there to be measured.

\subsection{Compatible and Incompatible \label{sbct5.2}}

Thus one way, perhaps the most essential way, quantum physics differs from
classical physics is that \emph{projectors representing different quantum
  properties need not commute}. We will say that the projectors $P$ and $Q$ are
\emph{compatible} provided $PQ=QP$, and \emph{incompatible} if $PQ\neq QP$.
Likewise a PDI $\{P_j\}$ and another PDI $\{Q_k\}$ are \emph{compatible} if
every projector in one commutes with every projector in the other:
$P_jQ_k=Q_kP_j$ for every $j$ and $k$. Otherwise they are \emph{incompatible}.
In the compatible case there is a \emph{common refinement} consisting of all
products of the form $P_jQ_k=Q_kP_j$, and every property in the event algebra
associated with $\{P_j\}$ or with $\{Q_k\}$ is also in the event algebra
associated with this refinement.
Hence a very central issue in quantum foundations, and also for quantum
information theory if one wants to use PDI's as sample spaces, is what to do
when quantum projectors do \emph{not} commute with each other. There have been
various approaches.

Von Neumann was well aware of this problem, and together with Birkhoff invented
\emph{quantum logic} \cite{BrvN36} to deal with it. In the case of a
spin-half particle, quantum logic says that ``$S_x=+1/2$ \AND\ $S_z=-1/2$'' is
the property represented by the zero operator; that is, it is meaningful, but
it is always false. This means its negation ``$S_x=-1/2$ \OR\ $S_z=+1/2$'' is
always true. Think about it: is that reasonable? If you continue to try and
apply ordinary logical reasoning in this situation, you will soon end up in
difficulty; see Sec.~4.6 of \cite{Grff02c} for details. To prevent paradoxes,
Birkhoff and von Neumann modified some of the rules of ordinary logic. Alas,
their quantum logic requires a revision of the rules of ordinary
(propositional) logic so radical that no one (known to me) has succeeded in
using it to think in a useful way about what is going on in the quantum world.
Maybe we physicists are just too stupid, and will have to wait for the day when
clever quantum robots with intelligence vastly superior to ours can use quantum
logic to resolve the quantum mysteries. But if they succeed, will they be able
to (or even want to) explain it to us?

A second approach to the incompatibility problem is employed in quantum
textbooks and is also widespread in the quantum foundations community. Instead
of talking about the quantum properties revealed by measurements, discussion is
limited to measurement \emph{outcomes}, the pointer positions that are part of
the macroscopic world where classical physics is an adequate approximation to
quantum physics, and noncommutation can be ignored for all practical purposes.
(More in Sec.~\ref{sct6} below.) I call this the ``black box'' approach
to quantum foundations. One starts with the \emph{preparation} of a microscopic
quantum state using a \emph{macroscopic} apparatus, and then a later
\emph{measurement} of the state using another \emph{macroscopic} apparatus, and
what lies in between---well, that is inside the black box, and we will
say as little as possible about it. A quantum $\ket{\psi}$? That is just a
symbolic way of representing the preparation procedure. A PDI $\{P_j\}$? That
is nothing but a mathematical tool for calculating the probabilities of
measurement outcomes. The black box approach has the advantage that it avoids
the problem of noncommuting quantum projectors. Its disadvantage is that it
provides no way of understanding in physical terms what is going on at the
microscopic level inside the box.

A third approach was popularized by Bell and his followers: replace the
\emph{noncommuting} Hilbert space projectors with \emph{commuting} hidden
variables. In essence, assume that in some way classical physics applies at the
microscopic level. But if, as I believe, noncommutation of projectors and PDI's
marks the frontier between classical and quantum physics, one should not be
surprised that an approach which is fundamentally classical---assumes a
classical sample space, as is evident from the way the mysterious symbol $\lm$
is employed in formulas---results in the famous Bell inequality that disagrees
with both quantum mechanical calculations and experimental results. (Nonlocal
influences can be ignored, since they do not exist; see \cite{Grff11}.)

\subsection{The Single Framework Rule\label{sbct5.3}}

The solution to the incompatibility problem that I favor can be viewed as a
lowbrow form of quantum logic, one that a physicist like me can actually make
use of. Its essential idea is that as long as one is dealing with a
\emph{single} PDI the rules of classical reasoning and classical probability
theory can be applied unaltered in the quantum domain. So let's do that. If two
PDI's are compatible, there is a PDI which is a common refinement. So let's use
it. But if two PDI's are incompatible, combining them will lead to nonsense. So
don't do it. These ideas have been worked out in considerable
detail in the \emph{consistent histories} (CH) interpretation of quantum
mechanics, where the prohibition against combining incompatible PDI's
is known as the \emph{single framework rule}. Here the term
\emph{framework} is used either for a PDI or the associated event algebra, and
the single framework rule prohibits combining incompatible PDI's.
The difference between CH and quantum logic can be illustrated using the
example ``$S_x=+1/2$ \AND\ $S_z=-1/2$'' discussed earlier. In quantum logic
this is meaningful but false, while in CH it is meaningless, neither true nor
false. The negation of a false statement is a true statement, so quantum logic
has to say something about it. But the negation of a meaningless statement is
equally meaningless, allowing CH to remain silent. See \cite{Grff14}
for more details.

In order to discuss the \emph{time development} of quantum systems a similar
approach can be used (the ``histories'' part of consistent histories). Once
again probabilities are assigned using PDI's as sample spaces, but in this case
on an extended Hilbert space of histories \cite{Ishm94}. In addition, in order
to assign probabilities to a family of histories (a PDI on the history sample
space) using an extension of the Born rule, it is necessary to impose certain
\emph{consistency} conditions (the ``consistent'' part of consistent
histories), if this family is to constitute an acceptable framework, so the
single framework rule is extended to incorporate the consistency conditions.
For a short introduction to the CH interpretation of quantum mechanics, see
\cite{Grff14b}. Various conceptual difficulties are discussed in \cite{Grff14},
whereas \cite{Grff13} gives a fairly thorough discussion of the ontology
(Hilbert subspaces as ``beables''). Finally, \cite{Grff02c} is a standard
reference with lots of details.

One aspect of the CH approach has raised a lot of objections, so it deserves a
comment. In a given situation it may be possible to describe what is going on
using various different but incompatible frameworks, so the question arises:
``What is the \emph{right} framework to use?'' The right answer is that this is
the wrong question to ask in the quantum domain. In classical mechanics the
state of a mechanical system at a particular instant of time can be exactly
specified by a single point in its phase space, the intersection of all
properties (sets of points) which are ``true'' at that instant. This is
consistent with the idea, which I have elsewhere called \emph{unicity}
(Sec.~27.3 of \cite{Grff02c}), that at every instant of time there is a single
unique ``state of the universe'' which, even if we do not know what it is,
determines all physical properties. What might be its quantum counterpart? A
``wavefunction of the universe''? If there really is something of that sort, it
is likely to be a horrible, uninterpretable superposition of different pointer
positions at the end of a measurement, or some other form of Schr\"odinger cat.
The corresponding projector will then not commute with properties that might
resemble something in the ordinary macroscopic world, and the single framework
rule will then prevent discussing the world of everyday experience. I do not
see any way in which a single quantum state could plausibly represent the
``true state of the world'', and I believe unicity must be
abandoned in the transition from classical to quantum physics.

In practice the choice of which framework to use will depend on the problem one
is interested in. Consider, for example, a situation in which a spin-half
particle is prepared in an eigenstate of $S_x$, say $S_x=+1/2$, before being
sent through a magnetic field-free region (so its spin direction will not
change) into an $S_z$ measuring device. The outcome of the measurement will be
either $S_z=+1/2$ or $S_z=-1/2$; let's assume the latter. This means we can say
that $S_z$ was $-1/2$ just before the measurement took place. But is it
possible that the particle had both $S_x=+1/2$ (because it was prepared in this
state) and $S_z = -1/2$ (the value measured later) \emph{at the same time},
just before the measurement was made? This makes no sense, as the properties are
incompatible. There is one framework in which at the intermediate time
$S_x=+1/2$, reflecting its earlier preparation, and a different, incompatible
framework in which at the intermediate time $S_z=-1/2$, reflecting the outcome
of the later measurement. These frameworks cannot be combined, and each has its
own uses. If we are concerned about whether $S_x$ was perturbed (say by a stray
magnetic field), then the $S_x$ framework is helpful, while if we want to
identify what the measurement measured, the $S_z$ framework is helpful. In
textbook quantum mechanics only the $S_x$ framework is employed. Nothing wrong
with that, except that one cannot discuss in what way the measurement measures
something, leaving the poor student rather confused.

This example suggests that the liberty to choose different frameworks is not as
dangerous as it might at first appear. A particular choice yields some type of
information, and a different choice may yield something different. By looking
at a coffee cup from above you can tell if it contains some coffee, while to
see if there is a crack in the bottom you need to look from below. The oddity
about the quantum world is not that different views, different frameworks, are
possible. Instead it is that certain frameworks cannot be combined into
a consistent quantum description, because they are incompatible. 
For another, less trivial, example of a case in which choosing alternative
frameworks proved useful, see the end of Sec.~\ref{sct7}.

\section{Quantum Information Theory I\label{sct6}}

Once a proper \emph{quantum} sample space, a PDI or framework, has been defined,
standard (Kolmogorov) probability theory can be used, and this means that the
whole machinery of classical (Shannon) probability theory can be imported,
unchanged, into the quantum domain. But the reasoning and the results are
restricted to this \emph{single framework}; in particular, they cannot be
combined with the analysis carried out in a separate, incompatible framework.
Probabilities associated with incompatible frameworks cannot be combined;
paying attention to this this eliminates a lot of well-known quantum paradoxes.
(See Chs.~19 to 25 of \cite{Grff02c}.)
 
In particular this provides a \emph{quantum} justification for all the usual
applications of classical information theory to macroscopic properties and
their time development. The reason is that from a quantum perspective the
classical mechanics of macroscopic objects can be discussed with quite adequate
precision using a single \emph{quasiclassical} quantum framework, in which
ordinary macroscopic properties are represented by enormous subspaces---a
dimension of $10$ raised to the power $10^{16}$ should be counted as relatively
small---whose projectors commute with one another for all practical purposes;
and quantum dynamics, which is intrinsically stochastic, is well 
approximated by deterministic classical dynamics. See \cite{GMHr93}; Chs.~7,
17, 18 of \cite{Omns99}; Ch.~26 of \cite{Grff02c}; and Sec.~4 of \cite{Grff13}.
Consequently, we can immediately claim that all of classical information
theory, all seventeen chapters of Cover and Thomas \cite{CvTh06}, or name your
favorite reference, are a valid part of \emph{quantum} information theory when
it is applied to macroscopic properties and processes. In this domain we
understand quite well what \emph{quantum} information is all about: its
probabilities refer to quasiclassical properties and processes, all the things
for which classical physics provides a satisfactory approximation to a more
exact quantum description.

(It is worth remarking, in passing, that using a quasiclassical framework
provides a solution to the infamous \emph{measurement problem} of quantum
foundations: what to do with a wavefunction which is a coherent superposition
of states in which the pointer points in two (or more) directions. While in the
CH approach there is nothing inherently wrong with such a thing, it can be
ignored if one wants to describe the usual macroscopic outcomes of laboratory
experiments. Use a quasiclassical framework, and the problems represented by
Schr\"odinger's cat are absent---and, by the single framework rule, they are
excluded from the description.)

In addition, Shannon's theory can be employed, unchanged, in situations in
which some or all of the properties being discussed are \emph{microscopic,
  quantum properties}, provided the discussion is restricted to a \emph{single
  framework}. This includes what I have elsewhere \cite{Grff15}
referred to as the \emph{second measurement problem}: inferring from the
measurement outcome (the pointer position) something about the earlier
microscopic state of the system being measured. It can be analyzed in a manner
which demonstrates that my colleagues who carry out experiments at accelerator
laboratories are not being foolish when they assert that a fast muon has
triggered their detector. The measurement apparatus is, in effect, an
information channel leading from microscopic quantum properties at the input to
macroscopic quantum properties (pointer positions) at the output.

\section{Quantum Information Theory II\label{sct7}}

Does this mean that \emph{all} problems of \emph{quantum} information can be
reduced to problems of \emph{classical} information? No, not at all, but it
does provide some insight into the nature of the additional problems which are
unique to quantum information, and what is needed to attack them. These
problems, and there are a vast number, all have to do with \emph{comparing}
(but not \emph{combining}!) situations involving \emph{incompatible
  frameworks}. But how can this be if a strict application of the single
framework rule is needed to avoid falling into nonsensical paradoxes? The
answer will emerge from considering some examples, starting with that of a noisy
quantum channel.

Consider a one-qubit memoryless quantum channel whose input and output is a
two-dimensional Hilbert space, the quantum analog of a classical one-bit
channel. The classical channel is characterized by two real parameters: the
probability that a 0 entering the channel will emerge as a 1, and the
probability that a 1 entering the channel will emerge as a 0. If both are zero,
the channel is perfect, noiseless. I like to visualize a perfect one-qubit
quantum channel as a pipe through which a spin-half particle is propelled in
such a way that its spin is left unchanged. If it enters with $S_x=+1/2$ it
exits with $S_x=+1/2$, if it enters with $S_z = -1/2$ it exits with $S_z =
-1/2$, and so forth. Of course, on any particular run the particle can only
have a well-defined spin angular momentum in a particular direction; e.g., it
can be prepared in such a state, and when it comes out only one component of
its spin angular momentum can be measured. So to \emph{test} whether the
channel is perfect it is necessary to carry out many repeated measurements.
This by itself is no different from a classical channel, where repeated
measurements are needed to estimate the probabilities of a bit flip when a
signal passes through the channel. However, in the quantum case the
probabilities that $S_z$ gets flipped, either from $+1/2$ to $-1/2$, or from
$-1/2$ to $+1/2$, can be very different from those for $S_x$, so repeated
measurements need to be carried out using different components of the spin
angular momentum. The single framework rule does not prohibit a discussion of
both $S_x$ and of $S_z$ \emph{provided} these refer to \emph{different} runs of
the experiment. There is no problem in supposing that in one run $S_z=-1/2$, on
the next run $S_x=+1/2$, and so forth. Of course, one has to assume that the
channel continues to behave in the same way, at least in a probabilistic sense,
during successive runs, but the same is true for a classical channel.

Suppose Joe has built what he claims is a perfect channel, but we want to test
it. This is straightforward for a 1-bit classical channel: send in a series of
0's and 1's, and see if what emerges from the channel is the same as what was
sent in. A one-qubit quantum channel is more complicated. If we test it using a
sequence of states in which $S_z=+1/2$ or $-1/2$, and what emerges is the same
as what went in, this is not sufficient, as it could very well be the case that
if one sends in $S_x=+1/2$ it will emerge with $S_x$ either $+1/2$ or $-1/2$ in
a completely random fashion, uncorrelated with the input. So we have to check
something in addition to $S_z$. Does this mean we have to carry out experiments
with $S_w=+1/2$ and $-1/2$ for \emph{every} possible spin component $w$? That
would take a lot of time, and is not necessary. It suffices to check both
$S_z=\pm 1/2$ and $S_x =\pm 1/2$. This result is far from obvious, and to
derive it one must use principles of quantum mechanics which have no classical
analog. Quantum information theorists need not fear unemployment; we will be
kept busy for a long time.

As another example, consider teleportation, often presented as an instance of
the mysterious and almost magical way in which quantum mechanics goes beyond
classical physics. A standard textbook presentation of a protocol to teleport
one qubit, e.g., Sec.~1.3.7 of \cite{NlCh00}, consists in applying unitary time
evolution to an initial quantum state, followed by a measurement which
collapses it. The measurement has four possible outcomes, and the result is
communicated from A to B through two uses of a perfect one-bit \emph{classical}
channel. The end result of the protocol is a quantum state transmitted
unchanged from A to B; in effect, a perfect one-qubit \emph{quantum} channel.
The student will certainly learn something by working through the formulas in
the textbook, but this is of limited value in developing an intuition about
microscopic quantum processes. My own approach \cite{Grff07} to understanding
teleportation employs two incompatible frameworks. One framework shows how
information about $S_x$ is transmitted from Alice to Bob with the assistance of
one use of the classical channel, and the other how $S_z$ information is
transmitted with the help of the other use of the classical channel. Similar
ideas (but without referring to frameworks) will be found in \cite{RnDR12} and
\cite{ClPn14}. This way of ``opening the black box'' should, I think, assist
students in gaining a better intuition for microscopic quantum processes, and I
hope it will become more widespread in the quantum information community, where
research, or at least its publication, is still dominated by the ``shut up and
calculate'' mentality encouraged by textbooks.

The preceding example could be easily dismissed in that it did not lead
(directly, at least) to any new results in quantum information: the original
teleportation protocol \cite{Bnao93} appeared fourteen years in advance of my
analysis. Hence it may be worth mentioning another example. A student and I
were trying to understand Shor's algorithm for factoring numbers, which ends
with a quantum Fourier transform followed by measurements of each of the qubits
in the standard basis $\ket{0}, \ket{1}$ basis ($\ket{z^+}, \ket{z^-}$ for a a
spin-half particle). We noted that if you suppose that the final measurement
reveals a property that the qubit possessed \emph{before} the measurement,
there is a way of looking at the problem that leads to an alternative and
simpler way to carry out the algorithm \cite{GrNi96}. Our perspective required
using a framework incompatible with that employed in the standard textbook
approach: unitary time development right up to the moment when measurement
``collapses'' the wavefunction---which, when done properly, leads to the same
final answer. I was pleased that Nielsen and Chuang mentioned our work
(Exercise 4.35 on p.~188, and see p.~246 of \cite{NlCh00}), but disappointed in
that they presented it as part of one more phenomenological principle, rather
than as a way of gaining insight by using measurements outcomes to infer
something about what happened earlier.

In my opinion, the discipline of quantum information could benefit from paying
attention to the developments in quantum foundations mentioned above. If you
open your favorite book on quantum information you will discover that
measurements are quite firmly imbedded in the discussion, and this in the
manner of other textbooks in which measurements do not actually measure
something, but instead enter as a primitive concept without further definition,
a rule for carrying out calculations which requires no real physical
understanding of processes at the microscopic quantum level. My guess is that
if quantum information texts were to provide a consistent discussion of
microscopic properties and processes, it could lead to some new and interesting
advances, and perhaps even some new insights into quantum foundations.

\section{Conclusion \label{sct8}}

Bell's question, ``Quantum information ... about what?'' can be given a quite
definite answer. It is about physical properties and processes, which in
quantum theory are represented by subspaces of the quantum Hilbert space, and
to which standard (Kolmogorov) probabilities can be assigned, using sample
spaces constructed from projective decompositions of the identity operator
(PDI's). The single framework rule of consistent histories forbids combining
incompatible PDI's or frameworks, resulting in a consistent theory not troubled
by unresolved quantum paradoxes. From a quantum perspective classical (Shannon)
information theory is the application of quantum information theory to the
domain of macroscopic properties and processes, where a single quasiclassical
quantum framework is sufficient for all practical purposes, and therefore
quantum incompatibilities can be ignored. But in addition, all the ideas of
classical information, and in particular its probabilistic formulation, can be
imported unchanged into the microscopic quantum domain, as long as one is
considering only a \emph{single} quantum framework.

That there are many distinct frameworks available in quantum theory, frameworks
which cannot be combined but can be compared, represents the new frontier of
information theory that is specifically \emph{quantum}, where classical ideas
no longer suffice. At this point new, and sometimes very difficult, problems
arise in the process of comparing (but not combining) different incompatible
quantum frameworks. They have no analogs in classical information theory, and
some of them are quite challenging. Progress in this domain might well benefit
were textbooks to abandon their outdated ``black box'' approach to quantum
theory, in which ``measurement'' is an undefined primitive and measurements do
not actually measure anything, but are simply a calculational tool to collapse
wavefunctions. It is past time to open the black box with tools that can
consistently handle noncommuting projectors. Consistent histories provides one
approach for doing this; if the reader can come up with something better, so
much the better.

\section*{Acknowledgments}

Major contributions to the consistent histories interpretation of quantum
mechanics have been made over the years by Roland Omn\`es, Murray Gell-Mann,
James Hartle, and, more recently, Richard Friedberg and Pierre Hohenberg. We
may not agree about everything, but I have certainly reaped great benefit from
conversations with and publications by these colleagues, and it is a pleasure
to thank them. I am also grateful for comments from three anonymous referees.

\end{document}